\documentclass[superscriptaddress,showpacs,amsmath,amssymb,aps,prb,twocolumn]{revtex4}
\usepackage[normal]{subfigure}
\usepackage{dcolumn}
\usepackage{bm}
\usepackage{pxfonts}

\usepackage[normalem]{ulem}
\usepackage{graphicx,color}
\usepackage{subfigure}
\usepackage{bbold}

\usepackage[colorlinks,citecolor=red,linkcolor=blue]{hyperref}
\usepackage[usenames,dvipsnames,svgnames]{xcolor}

\newcommand{\FF}{\mathcal{F}}

\newcommand{\ee}{\end{equation}}
\newcommand{\be}{\begin{equation}}
\newcommand{\MeijerG}[7]{G \begin{smallmatrix} #1 & #2 \\ #3 & #4 \end{smallmatrix} \left( \begin{smallmatrix} #5 \\ #6 \end{smallmatrix} \middle\vert #7 \right) }
\newmuskip\pFqmuskip

\newcommand*\pFq[6][8]{%
  \begingroup 
  \pFqmuskip=#1mu\relax
  \mathcode`\,=\string"8000
  \begingroup\lccode`\~=`\,
  \lowercase{\endgroup\let~}\pFqcomma
  {}_{#2}F_{#3}{\left[\genfrac..{0pt}{}{#4}{#5};#6\right]}%
  \endgroup
}
\newcommand{\pFqcomma}{\mskip\pFqmuskip}

\begin{document}
\title{Josephson coupling between superconducting islands on single and bilayer graphene}

\author{Francesco Mancarella}
\affiliation{Nordic Institute for Theoretical Physics
(NORDITA), SE-106 91 Stockholm, Sweden}
\affiliation{Department of Theoretical Physics, KTH Royal Institute of Technology, SE-106 91 Stockholm, Sweden}
\author{Jonas Fransson}
\affiliation{Department of Physics and Astronomy, Box 516, SE-721 51, Uppsala, Sweden}
\author{Alexander Balatsky}
\affiliation{Institute for Materials Science, Los Alamos, New Mexico 87545, USA.}
\affiliation{Nordic Institute for Theoretical Physics
(NORDITA), SE-106 91 Stockholm, Sweden}

\begin{abstract}
We study the Josephson coupling of superconducting (SC) islands through the surface of single-layer (SLG) and bilayer (BLG) graphene, as a function of distance between the grains, temperature, chemical potential and external (transverse) gate-voltage. For SLG, we provide a comparison with existing literature. The proximity effect is analyzed through a Matsubara Green function approach. This represents the first step in a discussion of the conditions for the onset of a granular superconductivity within the film, made possible by Josephson currents flowing between superconductors. To ensure phase coherence over the 2D sample, a random spatial distribution can be assumed for the SC islands on the SLG sheet (or intercalating the BLG sheets). The tunable gate-voltage-induced band gap of BLG affects the asymptotic decay of the Josephson coupling - distance characteristic for each pair of SC islands in the sample, which results in the end in a qualitatively strong field-dependence of the relation between Berezinskii-Kosterlitz-Thouless transition critical temperature and gate-voltage.
\end{abstract}
\date{\today}
\maketitle

\section{Introduction}\label{SecI}
The recent discovery of graphene \cite{Novoselov05, Zhang05}, a single sheet of carbon atoms, has naturally raised a question of superconductivity in this material \cite{Uchoa07} and its derivatives. This interest has been further stimulated by recent observation of proximity effect on graphene \cite{Heersche07, Du08}. We want to exhibit a potential for developing new superconducting devices starting with graphene as a basis material. Graphite, though not a superconductor in itself, can be made superconducting by intercalating certain dopants into its structure \cite{Dresselhaus02}. Among graphite-based materials which are known to
superconduct, the alkali metal-graphite intercalation compounds \cite{Koike80} have been widely investigated. The most easily fabricated among them is
the $\text{C}_8$K system \cite{Hannay65} which exhibits a transition temperature $T_c$=0.14 K \cite{Koike80}, while higher pressure allows to reach higher alkali metal concentration, such that the corresponding $T_c$ can increase up to 5 K in $\text{C}_2$Na \cite{Belash87}. On the other hand, lately Weller et al.
have shown that, at ambient conditions, the intercalated compounds $\text{C}_6$Yb and $\text{CaC}_6$ exhibit superconductivity with transition temperatures $T_c$=6.5 K and 11.5 K respectively \cite{Weller05}. 

It has been suggested \cite{Mazin10} that also Ca-intercalated bilayer graphene should be a superconductor with a critical temperature comparable to that of the 3D compound $\text{CaC}_6$ \cite{Belash02, Weller05, Emery05, Hannay65}. This prediction is based on a combination of linear augmented wave method for band structure calculations \cite{Blaha01} and density functional theory. 

Experiments based on metal-graphene hybrid composites have allowed the tuning of a proximity effect induced on graphene by superconducting nanoparticles deposited on top of it | decorating with tin clusters, separated by an average width much smaller than the mean free path and the superconducting coherence length \cite{Kessler10}, has been shown to induce a gate-tunable Berezinsky-Kosterlitz-Thouless (BKT) transition on micron-scaled exfoliated graphene samples.
More recently, a full electrical control of the superconductivity has been experimentally achieved for a centimeter-scale graphene sheet \cite{Allain12} on whose surface an array of tin nanoparticles is placed. Unlike the most common case of proximity coupling with the gate electrodes \cite{Heersche07}, in these two setups the proximity effect was generated by coupling the graphene surface to a 2D network of superconducting clusters \cite{Feigelman08}, so that the resulting hybrid systems globally behave as granular superconductors with universal transition threshold and Cooper pairs are localized in the insulating phase \cite{Allain12}. These experiments provide support of to the emergence of graphene as a backbone material for designing new superconductors. The proximity effect in graphene has inspired several possible applications, such as valley sensors \cite{Akhmerov07}, spin current filters \cite{Greenbaum07} and current switches \cite{Linder08, Lutchyn08}.
 
\begin{figure}[ht]
\begin{center}
\subfigure[]{
   \includegraphics[width=0.99\columnwidth] {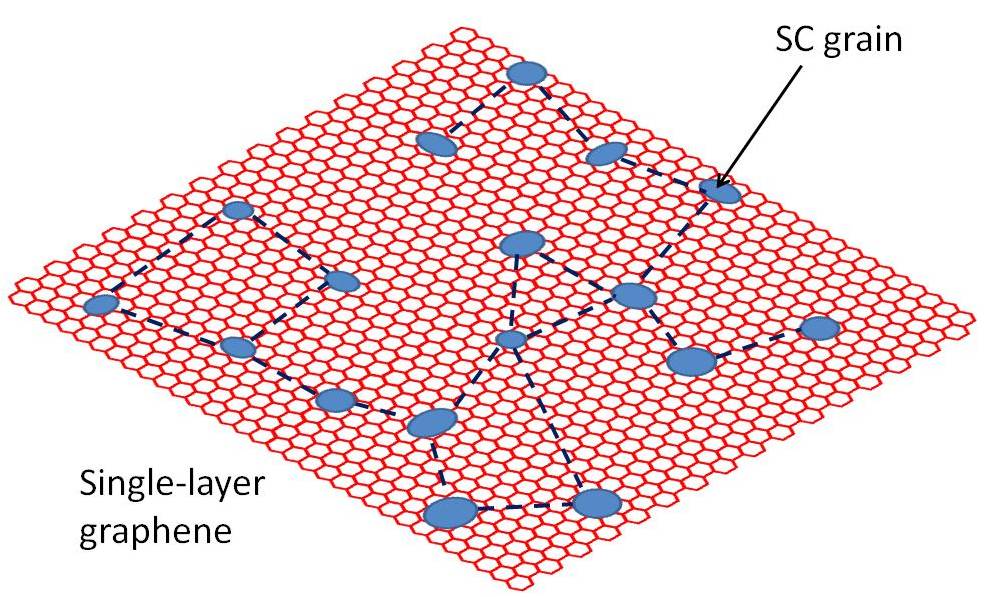}
   \label{grains_on_SLG}
 }

\subfigure[]{
   \includegraphics[width=0.99\columnwidth] {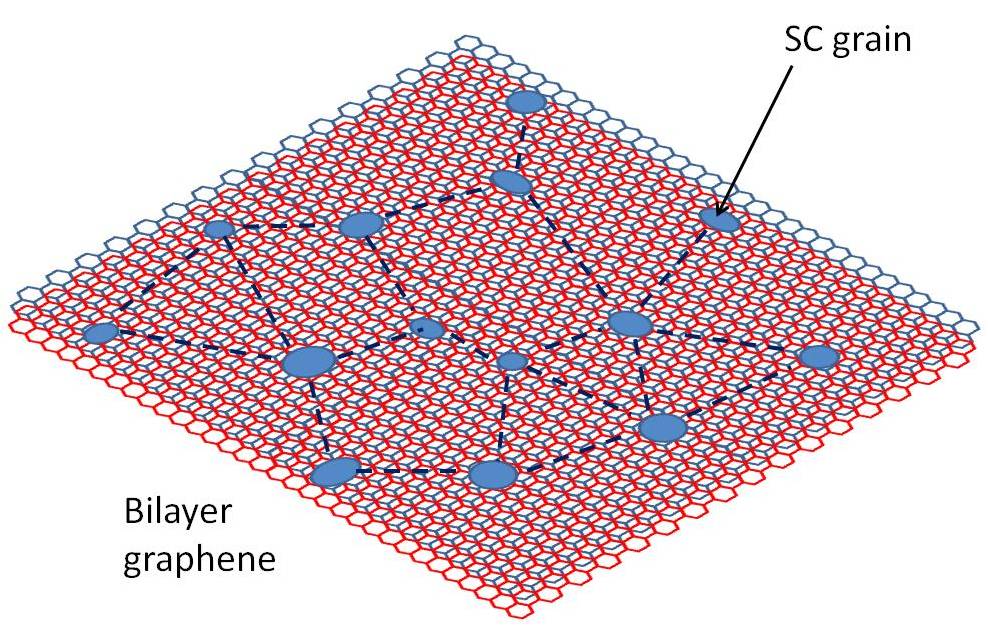}
   \label{grains_on_BLG}
 }
\end{center}
\caption{Cooper pair tunneling (dashed black lines) between superconducting grain impurities (blue spots), through \ref{grains_on_SLG} single-layer graphene and \ref{grains_on_BLG} bilayer graphene. In both cases, the entire carbon scaffold plays the role of Josephson junction connecting each couple of SC grains. For convenience of graphical representation, tunnelings are represented here only up to an arbitrary spatial range.}
\label{grains_on_G}
\end{figure}

Motivated by this scenario, we consider the proximity effect in single-layer graphene (SLG) and bilayer graphene (BLG) under various conditions of temperature, chemical potential, transverse electric field. In a complementary fashion w.r.t. the calculation of supercurrent in terms of Andreev reflection at the metal-superconductor interface \cite{Titov06}, we discuss the Josephson effect in the (opposite) dilute granular regime, i.e. when the distance $r$ between the SC adsorbates is much larger than their width $W$ and the superconducting coherence length $\xi$, such that the 2D nature of SLG/BLG has to be taken in account. In this regime the supercurrent is well described in terms of Cooper pairs tunneling through the SLG/BLG junction. The considered setups are schematically represented in Fig.~\ref{grains_on_G}, where the entire carbon scaffold behaves in principle as a junction connecting SC impurities. The Josephson current obtained in the SLG case at zero temperature, endorsing the result relevant to a clean undoped sample presented in Refs. \onlinecite{Gonzalez07,Gonzalez08}, reproduces also for the 2D case the coincidence between the supercurrent distance-decay for clean undoped SLG and disordered normal metals. This is just like it happens in the 1D case in the short junction regime \cite{Titov06, Tworzydlo06}. For both of these physical systems, the $\sim 1/r$ distance-decay of the supercurrent for 1D junctions \cite{Titov06} is replaced by a faster $\sim 1/r^3$ distance-decay for 2D Cooper pair propagation. We show that a finite doping turns a $\sim 1/r^3$ distance-decay into an asymptotic $\sim 1/r^4$ decay, for distances much beyond the critical length $\hbar v_F/\mu$. The temperature effect on the supercurrent is computed, and the asymptotic exponential decay $\sim r^{-2} e^{-2\pi r k_B T/ \hbar v_F}$ is derived for distances $r$ much larger than the thermal length $r_T$.  
The supercurrent across bilayer graphene is likewise discussed, resulting into a $\sim 1/r^2$ distance-decay at zero temperature, and an asymptotic exponential decay $\sim r^{-1} e^{-4\pi r/\lambda_T}$ much beyond the thermal length associated to the dispersion relation of BLG low-energy band. Finally we provide a comparison with former results in literature about SLG, while the asymptotic of the Josephson current through BLG is analytically derived.

We point out that decoration of graphene by means of extended superconducting mesoscopic grains is in principle not required to generate superconductivity. In fact, even an arrangement of local vibrational impurities is able to induce negative-$U$ centers in the carbon structure, as a result of the coupling between the local vibrational modes and the surrounding lattice \cite{Fransson13}. The fermionic state of the induced negative-$U$ center can be empty, singly or either doubly occupied with finite probability, and this suggests the formation of a local Cooper pair near each vibrational impurity \cite{Fransson13}. It should, therefore, be possible to effectively produce a local SC order parameter by simply placing pointwise vibrational impurities on top of the graphene sheet. 

The discussion presented in this work is twofold. First, the investigation of proximity effect \textit{per se}, and stimulated by practical proposals speficic to graphene junctions (which have been listed above). Second, the possibility of predicting features of the BKT phase transition for lattice-adsorbate graphene-based composites, as a function of the number density of lodged SC grain impurities, and the other parameters already mentioned. We retain a study of such a phase transition for later research. The paper is structured as follows: the supercurrent across a long Josephson junction is expressed in terms of the junction electron propagator in Sec.\,\ref{secII}; respectively for SLG and BLG, Sections\,\ref{secIII} and \ref{secIV} analyse the supercurrent radial dependence, considering the effect of doping and transverse gate-voltage, both at vanishing and finite temperature; in Sec.\,\ref{secV} the results of Sec.\,\ref{secIII} are confronted with the literature, while in Sec.\,\ref{secVI} we draw our conclusions; Sec.\,\ref{Acknowledgments} contains our  Acknowledgments, and the Appendix\,\ref{Appendix} specializes to our case a discussion of the asymptotic behavior of the Meijer functions. 

\section{Josephson current}\label{secII}
We begin by writing the expression for the critical Josephson current through a junction, in terms of the electron propagator of the junction and the parameters of the system. Letting $G$ denote the electron propagator in momentum-energy coordinates, $\FF^{-1(\bullet)}$, $\FF^{(\bullet)}$ the inverse and direct Fourier transform w.r.t. the set of variables "$\bullet$", and disregarding the constant factor $2ev_F^2W^2\left(\rho t^2W/v_F\right)^2$ ($W\ll\vert\vec{r}\vert$ is the width of the superconductors, $\rho$ is the normal density of electron states, $\rho t^2W/v_F$ is a dimensionless constant, $t$ the inter atomic hopping integral, and $v_F$ the Fermi velocity), the supercurrent $I(\vec{r})$ is given by \cite{Gonzalez08,Gonzalez07}:
\begin{align}
\FF^{-1(k)}[ i\, \text{Tr}\; G \stackrel{(\omega, k)}{*}\,G\;]_{\vert_{\omega=0}}=i\, \text{Tr}\;\tilde{G}(r,\omega) \stackrel{(\omega)}{*} \tilde{G}(r,\omega)_{\vert_{\omega=0}}.
\end{align}
Here, $\tilde{G}$ denotes the electron propagator in space-energy coordinates, the trace is performed by projecting over the singlet state of the Cooper pairs (see later sections for SLG and BLG cases), and $\stackrel{(\bullet)}{*}$ denotes the continuous/discrete convolution w.r.t. the set of variables "$\bullet$". Therefore the zero/finite temperature supercurrent is ($\hbar=1$)
\begin{subequations}
\begin{align}
I(\vec{r},T=0)\propto&
	 \,i\,\text{Tr}\;\,\int_{-i \infty}^{i \infty}\frac{d\omega_0}{2\pi}\;\tilde{G}(\vec{r},\omega_0) \cdot \tilde{G}(\vec{r},-\omega_0),
\label{zerotemperature}
\\
I(\vec{r},T)\propto&
	-k_B T\sum_{n=-\infty}^{\infty}\, \text{Tr}\;\tilde{G}(\vec{r},i\omega_n) \cdot \tilde{G}(\vec{r},-i\omega_n),
\label{finitetemperature}
\end{align}
\end{subequations}
Formula (\ref{finitetemperature}) reproduces formula (\ref{zerotemperature}) in the zero temperature limit.
We remark that this approach to the study of the Cooper pair propagation through SLG and BLG surfaces does not require any cutoff prescription.

\section{Single-layer Graphene}\label{secIII}
\subsection{Zero temperature}
For $T\rightarrow 0$, the calculation is automatically regularized by taking the zero temperature limit of the Matsubara summation. Now we write down the propagator for SLG in the desired coordinates. Here, and henceforth, we shall let $v_Fk\rightarrow k$ for a shorter notation.
\begin{align}
\tilde{G}(\vec{r},i\omega)=&
	\left[\int d^2k\;e^{ik\cdot r}\frac{-i\omega}{\omega^2+k^2}\right]\,\mathbb{1}
	-\vec{\sigma}\cdot \left[\int d^2k\;e^{ik\cdot r}\frac{\vec{k}}{\omega^2+k^2}\right]
\nonumber\\=&
	-2\pi i \omega K_0(\sqrt{\omega^2}\,r)\,\mathbb{1}-2\pi i\sqrt{\omega^2}\,K_1(\sqrt{\omega^2}\,r)\, \vec{\sigma}\cdot \hat{r},
\label{slgpropagator}
\end{align}
where $K_\nu(z)$ is the modified Bessel function of the second kind (with argument written in units of $1/v_F$), $\mathbb{1}$ is the $2\times 2$ identity matrix in each sublattice space. We calculate the trace by starting with the following formal notation for the propagator above (with obvious shorthand notations $g_1$,$g_2$, and by denoting $\tilde{g_i}(\omega)\equiv g_i(-\omega), i=1,2$):
\be 
\tilde{G}(\vec{r},i\omega)=g_1\,\mathbb{1}+g_2\, \vec{\sigma}\cdot \hat{r}.
\ee
The initial and final singlet states ($\epsilon_{\alpha\beta}$ antisymmetric tensor of rank 2) of the Cooper pairs requires:
\begin{align}
\text{Tr}_{(sublattice)}&\;\tilde{G}(\vec{r},i\omega_n) \cdot \tilde{G}(\vec{r},-i\omega_n)
\nonumber\\\equiv&
	\epsilon_{\alpha\beta}(g_1\,\mathbb{1}_{\alpha\mu}+g_2\, 
	(\vec{\sigma}\cdot \hat{r})_{\alpha\mu})(\tilde{g_1}\,\mathbb{1}_{\beta\nu}+\tilde{g_2}\, 
	(\vec{\sigma}\cdot \hat{r})_{\beta\nu})\epsilon_{\mu\nu}
\nonumber\\=&
	2(g_1\tilde{g_1}+g_2\tilde{g_2}),
\end{align} 
therefore the \text{Tr} symbol acts as a simple overall 2 factor. As a consequence, the integration along the imaginary $i\omega_0$-axis gives
\begin{align}
I(\vec{r})\propto&
	i\, \text{Tr}\;\,\int_{-i \infty}^{i \infty}\frac{d (i\omega_0)}{2\pi}\;\tilde{G}(\vec{r},i\omega_0) \cdot \tilde{G}(\vec{r},-i\omega_0)
\nonumber\\=&
	\frac{4}{r^3}\,\int_{0}^{\infty}\frac{d x}{2\pi}\,x^2\,\left[-4\pi^2(K_0(x))^2+4\pi^2(K_1(x))^2\right]
\nonumber\\=&
	\frac{\pi^3/2}{r^3},
\label{monolayeru0current}
\end{align}
which agrees with the distance decay of the supercurrent described in Refs. \onlinecite{Gonzalez08,Gonzalez07}, although differing from that by an overall constant factor.

\subsection{Finite temperature}
For finite temperatures we use Eq. (\ref{finitetemperature}), which yields
\begin{align}
I(\vec{r},T)\propto&
	16\,\pi^4 \, (k_BT)^3 \sum_{n=0}^{\infty}\,(2n+1)^2
\nonumber\\&\times
	\left[-(K_0((2n+1)\pi x))^2+(K_1((2n+1)\pi x))^2\right] \;,
\label{slgfiniteTcurrent}
\end{align}
where $x\equiv r k_B T\gg 1$ for distances much larger than the thermal length $r_T$. For large distances $r\gg r_T$, the supercurrent can be expressed in analytical form as
\begin{align} 
I(\vec{r},T)\approx&
	\frac{ 8\, \pi^3\,k_B T}{r^2}\,e^{-2\pi x}, \quad r \gg r_T\text{ (i.e. }x \gg 1\text{)}.
\label{finiteTSLGdecay}
\end{align}
The asymptotic decay (far beyond the thermal length) is exponential, which is different from the quintic decay in Ref. \onlinecite{Gonzalez08}, Eq. (20) (see discussion in Appendix \ref{Appendix}).

\begin{figure}[t]
\vspace{0.4cm}
\centerline{
\includegraphics[width=0.99\columnwidth]{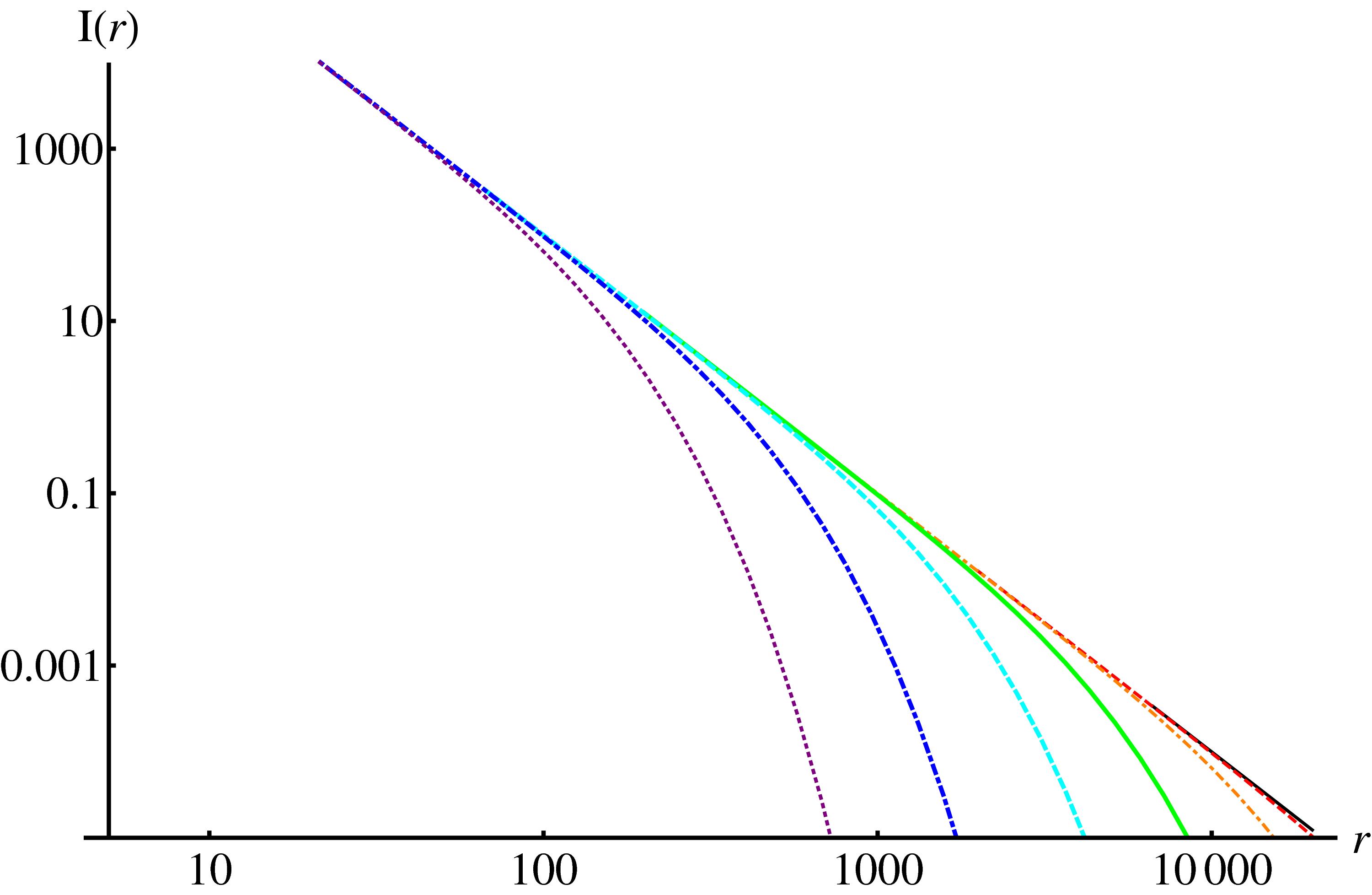}}
\caption{Radial dependence of the supercurrent $I(r)$ for graphene for different temperatures, from black (top) to purple (bottom) $T=[$ 0.140 (solid black), 0.43 (dashed red), 1.4 (dash-dot orange), 4.3 (thick solid green), 14 (thick dashed cyan), 43 (thick dash-dot blue), 140 (thick dotted purple)$]$  K. Here, $I(r)$ is given in units of $(\rho t^2 W/v_F)^2 \times$ 11.4 nA $\times$ $W^2[\text{nm}^2]=24.7\, \mu\text{A}$, where the dimensionless combination $\rho t^2 W/v_F$ is the relative conductance of each junction, $W[\text{nm}]$ the junction width expressed in nanometers, the Fermi velocity has been approximated as $v_F=10^6$ m/s, and the numerical value refers to the case of tin impurities of width $W=50$ nm.}
\label{arcobaleno}
\end{figure}

For arbitrary distance $r$, we show the behavior of $I(r)$ in Fig.\ref{arcobaleno} , obtained by numerical evaluation the sum (\ref{slgfiniteTcurrent}) for different temperatures. The knees of curves sit about respective crossover-distances of the order of the critical thermal length $r_T=\hbar v_F/k_B T$ for each temperature. Beyond this critical length $r_T$, the cubic distance-decay (\ref{monolayeru0current}) gradually veers towards the exponential decay (\ref{finiteTSLGdecay}).

\subsection{Finite doping}
Doping of graphene generates a shift of the Fermi level from the neutrality point. We can introduce this shift in our approach by adding a finite chemical potential $\mu$.  The corresponding Matsubara propagator for each Dirac electron is accordingly modified as follows
\begin{subequations}
\begin{align}
\tilde{G}^\mu(\vec{r},i\omega_n)=&
	\left[\int \frac{d^2k\;e^{ik\cdot r}}{i \omega_n-(\vec{\sigma}\cdot k-\mu)}\right]=\tilde{G}^{(\mu=0)}(\vec{r},i\omega')\,, 
\label{finitemupropagator}
\\
\tilde{G}^\mu(\vec{r},-i\omega_n)=&
	\left[\int\frac{ d^2k\;e^{ik\cdot r}}{-i \omega_n-(\vec{\sigma}\cdot k-\mu)}\right]=\tilde{G}^{(\mu=0)}(\vec{r},i\omega'')\,, 
\end{align}
\end{subequations}
where we introduce the notation $\omega'\equiv \omega_n-i \mu$, $\omega''\equiv -\omega_n-i \mu$. For any value of $\mu$, the square root $\sqrt{\omega'\,^2}=\pm \omega',\;\sqrt{\omega''\,^2}=\mp \omega''$; upper/lower signs refer to positive/negative $\omega_n$. Proceeding as in Eq. (\ref{monolayeru0current}), the supercurrent through doped graphene at zero temperature can be written in the following way ($\omega_0$ is the continuous variable version, for $T=0$, of the discrete variable $\omega_n$; $\omega'_0\equiv \omega_0-i \mu$\;, $\omega''_0\equiv -\omega_0-i \mu$ ):

\begin{figure}[t]
\centerline{
\includegraphics[width=0.99\columnwidth]{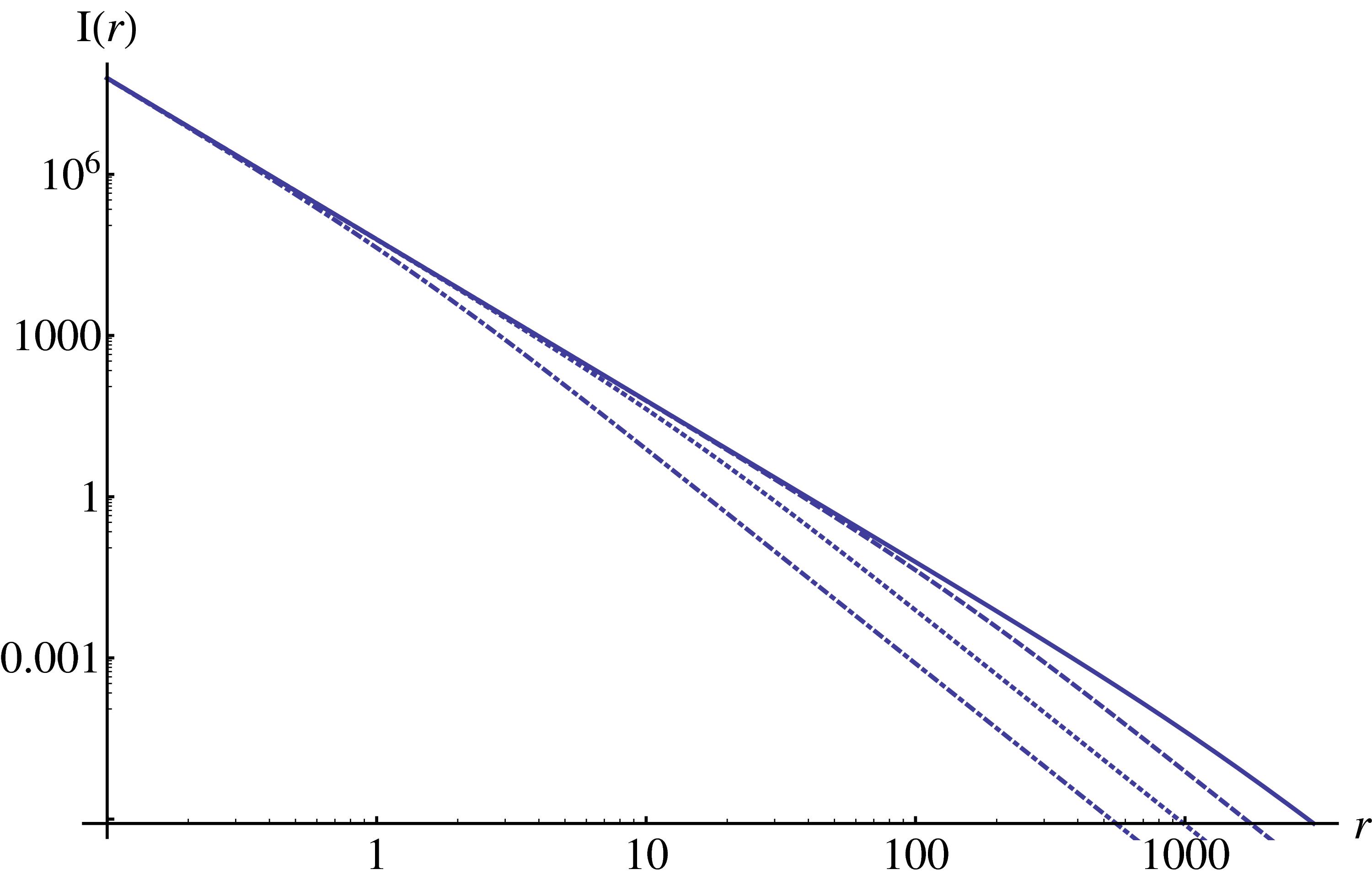}}
\caption{Radial dependence of the zero temperature supercurrent $I(r)$ for different chemical potentials, from top to bottom $\mu=[1 \text{ (solid)},\ 10 \text{ (dashed)},\ 100 \text{ (dotted)},\ 1000 \text{ (dash-dot)}]\times10^{-3}\;\hbar v_F/\text{nm} \approx 4.136$ meV. $I(r)$ is given in units of $(\rho t^2 W/v_F)^2 \times$ 80.5 nA $\times$ $W^2[\text{nm}^2]=174.5 \,\mu\text{A}$, where tin impurities of width $W=50$ nm have been considered. Other parameters are as in Fig. \ref{arcobaleno}.}
\label{figSLG_finitemu_zeroT}
\end{figure}

\begin{align} 
I^\mu(\vec{r})\propto&
	4\pi \int_{-\infty}^{\infty} d\omega_0\;\left\{ \left[\omega'_0\; K_0(\pm \omega'_0 r)\right]\left[\omega''_0\; K_0(\mp \omega''_0 r)\right]\right.
\nonumber\\&
	+\left.\left[\pm\omega'_0\; K_1(\pm \omega'_0 r)\right]\left[\mp\omega''_0\; K_1(\mp \omega''_0 r)\right]\right\}
\nonumber\\=&
	-8\pi\int_0^\infty d\omega_0\left\{K_0(r(\omega_0-i\mu))\,K_0(r(\omega_0+i\mu))\right.
\nonumber\\&
	\left.-K_1(r(\omega_0-i\mu))\,K_1(r(\omega_0+i\mu))\,\right\}
	(\omega_0^2+\mu^2)
	.
\end{align}
Using the transformation properties under conjugacy $K_{\overline{\nu}}(\overline{z})=\overline{K_\nu(z)}$, the above expression can be rewritten as
\begin{subequations}
\begin{align}
I^\mu(\vec{r})\propto&
	-\frac{8\pi}{r^3}f(\mu\,r),
\label{formula11a}	
\\
f(y)\equiv&
	\int_0^\infty dx\;(x^2+y^2)\left( \,\vert K_0(x-i y)\vert^2-\vert K_1(x-i y)\vert^2 \,\right)\,.
\end{align}
\end{subequations}
Noticing that $f(0)=-\pi^2/16$, it is clear that the result for $\mu=0$, Eq. (\ref{monolayeru0current}), is correctly reproduced. Asymptotically we have $f(y)=\pi y^{-1}/4+o(y^{-1})$. For finite $\mu$ and in the regime of large distances $r\gg \hbar v_F \mu^{-1}$, the supercurrent is therefore rewritten as 
\begin{align}
I^\mu(\vec{r})\approx&
	-\frac{2\pi^2}{\mu\,r^4},\quad r\gg \hbar v_F \mu^{-1}\,.
\end{align}
This result is qualitatively different from the inverse quadratic decay of the supercurrent derived in Refs. \onlinecite{Gonzalez08,Gonzalez07} under the similar assumptions. In fact, the decay described here shows a different asymptotic power-law, which does not include oscillations with the distance $r$. The crossover between the near-distance cubic decay and the long-distance quartic decay is indicated in Fig.\ref{figSLG_finitemu_zeroT}, where the set of curves corresponds to a wide range for the possible values of the chemical potential $\mu$. The knees of the four curves occur at respective crossover-distances of roughly, from top to bottom, $[1000,\ 100,\ 10,\ 1]$ nm.

\begin{figure}[t]
\begin{center}
\includegraphics[width=0.99\columnwidth]{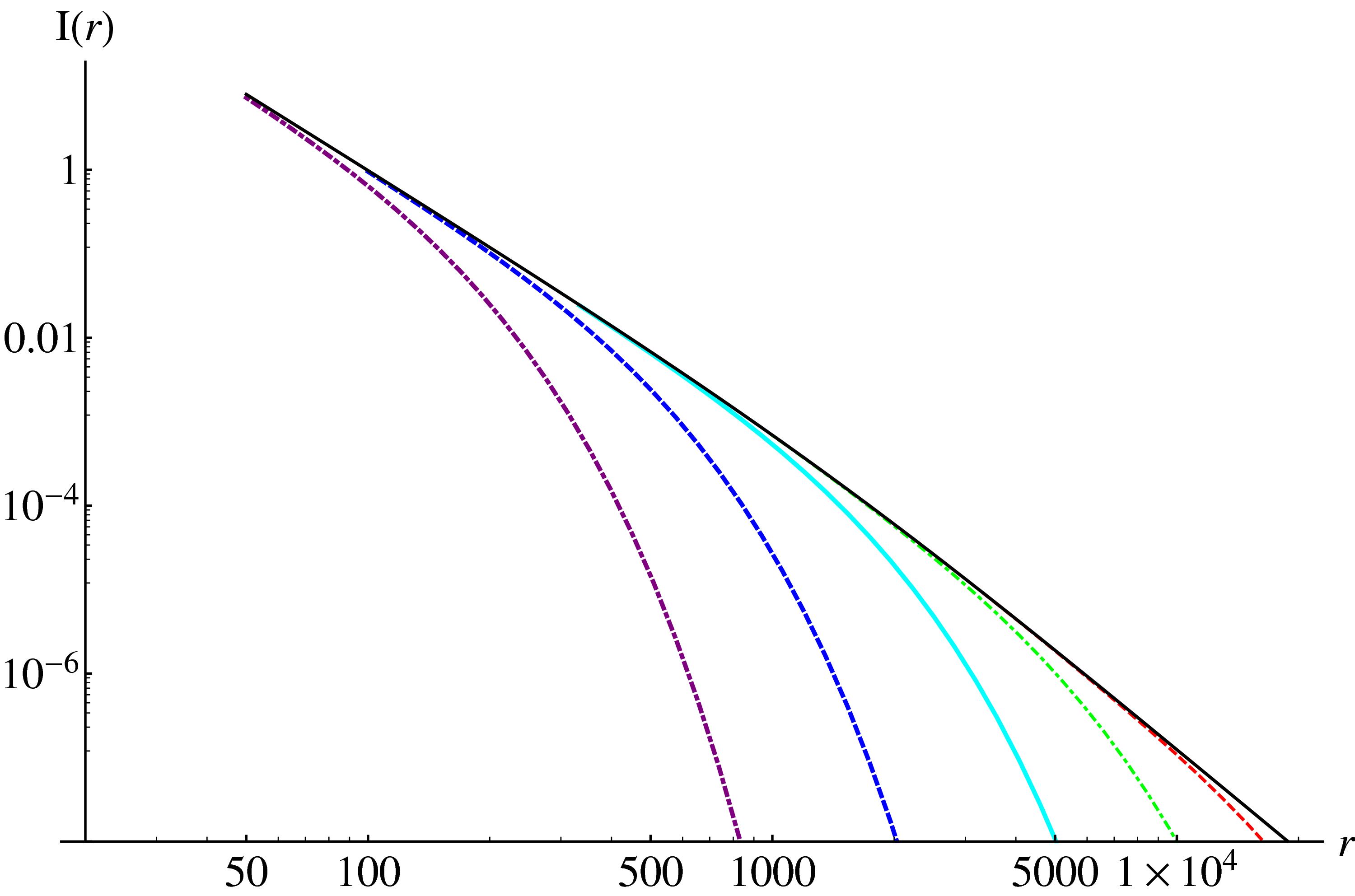}
\end{center}
\caption{Radial dependence of the supercurrent $I(r)$ through graphene for different temperatures, from top to bottom $T=[$0 (solid black),\ 1.4 (dashed red),\ 4.3 (dash-dot green),\ 14 (thick solid cyan),\ 43 (thick dashed blue),\ 140 (thick dash-dot purple)$]$ K, at finite chemical potential $\mu=10^{-3}\, \hbar v_F/\text{nm}\approx 4.136$ meV. The top black curve shows well the zero-temperature crossover from cubic to quartic distance-decay already presented in Fig. \ref{figSLG_finitemu_zeroT}. $I(r)$ in units of $(\rho t^2 W/v_F)^2 \times$ 4.96 nA $\times$ $W^2[\text{nm}^2]=10.8 \,\mu\text{A}$, where tin impurities of width $W=50$ nm have been considered. Other parameters are as in Fig. \ref{arcobaleno}.}
\label{finitemuarcobaleno}
\end{figure}

For finite temperature, we use Eqs. (\ref{finitetemperature}) and (\ref{finitemupropagator}) to obtain the supercurrent decay, here, given by
\begin{align} 
I^\mu(\vec{r})\propto&
	-16\, \pi^4 \,(k_B T)^3\,\sum_{n=0}^{\infty} \left\{\left((2n+1)^2+\left(\frac{y}{\pi x}\right)^2\right)\right.
\nonumber\\\times&
	\left.\left[\vert K_0\Bigl((2n+1)\pi x-iy\Bigr)\vert^2-\vert K_1\Bigl((2n+1)\pi x-iy\Bigr)\vert^2\right]\right\}\,,
\end{align}
where $x \equiv r k_B T$ and $y \equiv r \mu$. We plot the characteristics of $I^\mu(\vec{r})$ in Fig. \ref{finitemuarcobaleno}. We notice from the plots that the radial dependence of the supercurrent for finite temperature and doping is similar to the one we obtain for zero doping, c.f. Fig. \ref{arcobaleno}. Far beyond the thermal length (i.e. for $x\gg1/4\pi)$, the asymptotic expansion for the modified Bessel function gives the result
\begin{align}
I^\mu(\vec{r})\approx&
	\frac{ 8\, \pi^3\,k_B T}{r^2\;\sqrt{1+\left(\frac{\mu}{\pi k_B T}\right)^2}}\;e^{-2\pi x}\,, \quad r\gg r_T/(4\pi).
\end{align}
The expression reduces to the one given in Eq. (\ref{finiteTSLGdecay}) in the non-degenerate limit $\mu\ll \pi k_B T$, and to the expression
\begin{align}
I^\mu(\vec{r})\approx&
	\frac{ 8\, \pi^4\,(k_B T)^2}{\mu\;r^2}\;e^{-2\pi x}\,, \quad r\gg r_T/(4\pi)\,\gg (4\mu)^{-1}
\end{align}
in the strongly-degenerate limit $\mu\gg \pi k_B T$.
 
\section{Bilayer graphene}\label{secIV}
We consider bilayer graphene made out of two coupled graphene sheets with inequivalent sites A1,B1 and A2,B2 respectively, on the bottom and top graphene sheets, arranged according in Bernal stacking (A2-B1). We limit ourselves to the low-energy approximation in which our model reduces to a $2\times2$ matrix formulation. We denote the effective Hamiltonian by $\hat{H}_2$, around the two valleys K ($\xi=+1$) and K' ($\xi=-1$). In addition, we employ the two simplifying approximations:
\begin{enumerate}
\item		trigonal warping $v_3=0$,
\item		$v_Fk/\gamma_1\ll 1$ in the significant domain of integration over momenta.
\end{enumerate}

Under those assumptions we obtain
\begin{subequations}
\begin{align}
\hat{H}_{2 \xi}(\vec{k})=&
	\vec{\sigma}\cdot \vec{A}_\xi(\vec{k}),
\\
\vec{A}_\xi(\vec{k})\equiv&
	\{-\frac{k^2}{2m}\cos(2 \theta_{\vec{k}}),
	-\frac{k^2}{2m}\sin(2 \theta_{\vec{k}}),
	-\xi \frac{u}{2}\},
\end{align}
\end{subequations}
where $m\approx0.017$ meV/$c^2$ denotes the effective particle mass for low-energy bands of bilayer graphene \cite{McCann2013, McCann07}, and $u$ the magnitude of the voltage associated to the electric field transverse to the BLG plane. The corresponding Green function in momentum space can be written
\begin{align} 
G_{BLG}(\vec{k},i\omega)=&
	\frac{i\omega+\hat{H}_{2 \xi}(\vec{k})}{-\omega^2-(\hat{H}_{2 \xi}(\vec{k}))^2}
	=\frac{-i\omega-\sigma\cdot \vec{A}_\xi(\vec{k})}{\omega^2+\vert\vec{A}_\xi(\vec{k})\vert^2},
\end{align}
from which we derive the real space GF
\begin{align}
\tilde{G}_{BLG}(\vec{r},i\omega)=&
	\int d^2k
		\,e^{i k r \cos \theta}
		\biggl(\omega^2+\biggl(\frac{k^2}{2m}\biggr)^2+\frac{u^2}{4}\biggr)^{-1}
		\Bigl(
			-i\omega\mathbb{1}
\nonumber\\&
			-\vec{\sigma}\cdot
				\{-\frac{k^2}{2m}\cos(2 \theta_{\vec{k}}),\ -\frac{k^2}{2m}\sin(2 \theta_{\vec{k}}),\ -\xi \frac{u}{2}\}
		\Bigr)
\nonumber\\=&
	-2\pi\left\{\left(i\,\omega\,\mathbb{1}+\frac{\xi\,u}{2}\sigma_z\right)\;f_1+(\vec{\sigma} \cdot \hat{r})\; f_2 \right\},
\label{blgrealspacepropagator}
\end{align}
where 
\begin{subequations}
\begin{align}
f_1=&
	\sqrt{\frac{m^2}{u^2 + 4\omega^2}}
	\MeijerG{3}{0}{0}{4}{-}{0,\frac{1}{2},\frac{1}{2},0}{\frac{1}{256} m^2 r^4 (u^2 + 4 \omega^2)},
\label{effe1}
\\
f_2=&
	\frac{m}{2}
	\MeijerG{3}{0}{0}{4}{-}{0,\frac{1}{2},1,-\frac{1}{2}}{\frac{1}{256} m^2 r^4 (u^2 + 4 \omega^2)}.
\label{effe2}
\end{align}
\end{subequations}
Here, we have used the traditional notation
\begin{align}
G_{p,q}^{\,m,n} \!\left( \left. \begin{matrix} a_1, \dots, a_p \\ b_1, \dots, b_q \end{matrix} \; \right| \, z \right)=&
	G_{p,q}^{\,m,n} \!\left( \left. \begin{matrix} \mathbf{a_p} \\ \mathbf{b_q} \end{matrix} \; \right| \,  z \right) 
\end{align}
for the Meijer function $G$. Below, we present the details for the components $f_1$ and $f_2$ of the bilayer graphene matrix propagator.

\subsection{Zero temperature}
The supercurrent depends on the product $\tilde{G}_{BLG}(\vec{r},i\omega_n) \cdot \tilde{G}_{BLG}(\vec{r},-i\omega_n)$. By tracing over the singlet state Cooper pairs of the BLG we obtain
\begin{align}
\text{Tr}_{(sublattice)}\tilde{G}_{BLG}(\vec{r},i\omega_n)\cdot\tilde{G}_{BLG}(\vec{r},-i\omega_n)&
\nonumber\\&\hspace{-4cm}=
	\epsilon_{\alpha\beta}(g_1\,\mathbb{1}_{\alpha\mu}+g_2\, (\vec{\sigma}\cdot \hat{r})_{\alpha\mu}+g_3 \sigma^z_{\alpha\mu})
\nonumber\\&\hspace{-3.7cm}\times
	(\tilde{g_1}\,\mathbb{1}_{\beta\nu}+\tilde{g_2}\, (\vec{\sigma}\cdot \hat{r})_{\beta\nu}+\tilde{g_3} \sigma^z_{\beta\nu})\epsilon_{\mu\nu}
\nonumber\\&\hspace{-4cm}=
	2(g_1\tilde{g_1}+g_2\tilde{g_2}+g_3\tilde{g_3}).
\end{align}
As in the SLG, the Tr effectively generates a factor 2. Inserting this result into the zero temperature current, Eq. (\ref{zerotemperature}), in absence of external field ($u=0$), we obtain
\begin{align}
I(\vec{r})\vert_{T=0}\propto&
	i\text{Tr}
	\int_{-i \infty}^{i \infty}\frac{d (i\omega_0)}{2\pi}\tilde{G}_{BLG}^{u=0}(\vec{r},i\omega_0)\cdot\tilde{G}_{BLG}^{u=0}(\vec{r},-i\omega_0)
\nonumber\\=&
	-\frac{8\pi^2m}{r^2}
\label{bilayeru0current}
\end{align}
In the following discussion we shall derive this quadratic decay of the supercurrent displayed in the last line of Eq. (\ref{bilayeru0current}). This qualitatively different characteristics compared to the cubic decay for SLG is striking and we return to the physical reason for this below.

First, the real space GFs expressed in Eq. (\ref{blgrealspacepropagator}), can be rewritten as follows (denoting $\vec{s}\equiv r\vec{k}$, $\Omega\equiv 2 m r^2\omega$)
\begin{align}
\tilde{G}_{BLG}(\vec{r},i\omega)=&
	-2im\Omega\left[\int d^2s \;e^{is\cos\theta}\frac{1}{\Omega^2+s^2+(umr^2)^2} \right]\mathbb{1}
\nonumber\\&
	-2m\vec{\sigma}\cdot\left[\int d^2s\; e^{is\cos\theta}\frac{\left(\begin{array}{c}-s^2\cos(2\theta_k)\\-s^2\sin(2\theta_k)\\-\xi u m r^2\end{array}\right)}{\Omega^2+s^2+(umr^2)^2} \right]
\nonumber\\=&
	m\left[g_1(\Omega,umr^2)\mathbb{1}+\vec{g}_2(\Omega,umr^2)\cdot \vec{\sigma}\right],
\end{align}
in obvious notation. Inserting this expression into the zero temperature current, Eq. (\ref{zerotemperature}), we obtain
\begin{subequations}
\begin{align}
I(\vec{r})\propto\,&
	m^2\int_{-\infty}^{\infty}d\omega \left[g_1^2(\Omega,umr^2)+g_2^2(\Omega,umr^2)\right]
\nonumber\\=\,&
	\frac{m}{r^2}g_3(umr^2),
\label{blgargument}
\\
g_3(x)\equiv\,&
	\frac{1}{2}\int_{-\infty}^{\infty}d\Omega\,\left[g_1^2(\Omega,x)+g_2^2(\Omega,x)\right],
\end{align}
\end{subequations}
hence, demonstrating the quadratic decay of $I(\vec{r})$ when no external field is applied. Physically this means that the low-energy parabolic dispersion relation of BLG automatically results in a quadratic radial decay of the supercurrent.

For SLG instead, the same simple scaling argument applied to the real space propagator in the $\mu=0$ case gives
\be 
\tilde{G}(\vec{r},i\omega)=\omega\left[g_4(\gamma)\,\mathbb{1}+\vec{g}_5(\gamma)\cdot\vec{\sigma}\right]\,,
\ee	
where $g_4$ and $g_5$ are suitable functions of $\gamma\equiv r\omega$, and the application of (\ref{zerotemperature}) yields
\be
I(\vec{r})\propto \int_{-\infty}^{\infty}d\omega\;\omega^2\,\left[g_4(\gamma)^2+\vec{g}_5(\gamma)^2\right]\,\propto\,\frac{1}{r^3},
\ee
so that the cubic radial decay is a consequence of the linear low-energy linear dispersion in SLG.

\subsection{Finite temperature}
\label{subsectBLGfiniteT}
We proceed by investigating the properties of the supercurrent for finite temperature. As we cannot analytically sum over the Matsubara frequencies for arbitrary distance, we approximate the supercurrent between two SC islands by truncating the sum (\ref{finitetemperature}).

\begin{figure}[t]
\begin{center}
\includegraphics[width=0.99\columnwidth]{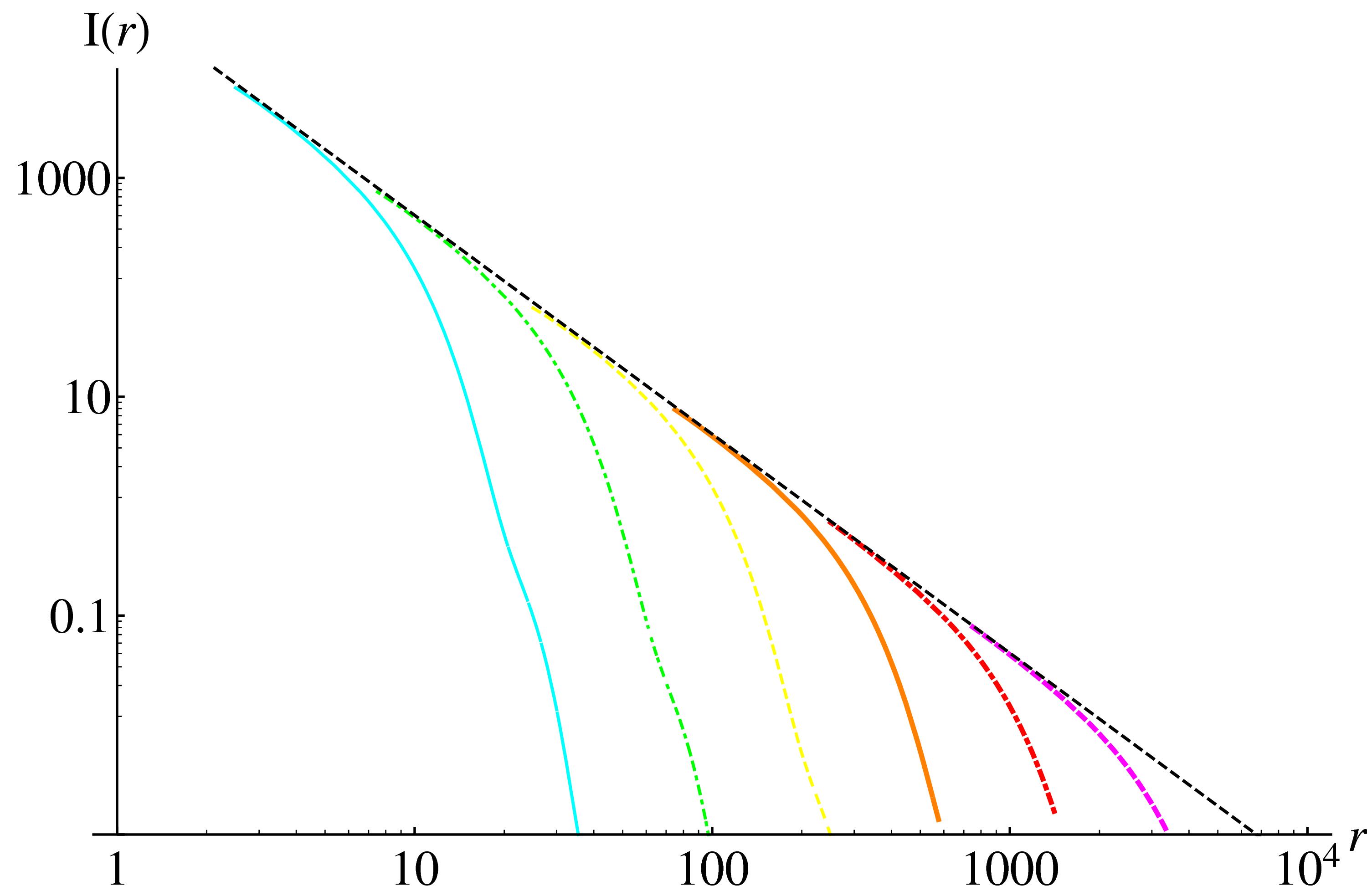}
\end{center}
\caption{Radial dependence of the supercurrent $I(r)$ for bilayer graphene for different temperatures, from top to bottom $T=[$ 0 (thin dashed black), 0.003 (thick dashed magenta), 0.03 (thick dash-dot red), 0.3 (thick solid orange), 3 (dashed yellow), 30 (dash-dot green), 300 (solid cyan)$]$ K. The top black thin dashed line shows well the zero-temperature quadratic distance-decay already presented in Eq. (\ref{bilayeru0current}). $I(r)$ in units of $(\rho t^2 W/v_F)^2 \times$ 160 nA $\times$ $W^2[\text{nm}^2]=8.88\, \mu\text{A}$, where tin impurities of width $W=20$ nm have been considered. Other parameters are as in Fig. \ref{arcobaleno}.
}
\label{BLGarcobaleno}
\end{figure}

For intermediate distances, the supercurrent can be written as
\begin{align} 
I(\vec{r},T)\propto &
	-4 k_BT \sum_{n=0}^\infty \tilde{G}_{BLG}(\vec{r},i\omega_n)\,\tilde{G}_{BLG}(\vec{r},-i\omega_n)
\nonumber\\=&
	-16\pi^2k_BT\sum_{n=0}^{\infty}\left[(\omega_n^2+\frac{u^2}{4})f_1^2(\omega_n)+f_2^2(\omega_n)\right].
\label{blgfinitetemperature}
\end{align}
In Fig.\,\ref{BLGarcobaleno} we plot the radial dependence of $I(r)$, which is obtained by calculating the series in Eq. (\ref{blgfinitetemperature}) numerically (notations in (\ref{effe1},\ref{effe2}) are used).

For large distances, $m r^2 k_B T\, \hbar^{-2} \gg 1$, the evaluation of (\ref{blgfinitetemperature}) can be traced back to the asymptotic expansion of the Meijer function \cite{Fields72}. This expansion is not automatically workable by commonly used computer algebra systems and, therefore, we adapt the approach discussed in Ref. \onlinecite{Fields72}, which is briefly presented in Appendix \ref{Appendix}.

First, we write the asymptotic behaviors
\be \MeijerG{3}{0}{0}{4}{-}{0,\frac{1}{2},\frac{1}{2},0}{z} \sim 
c_1\, z^{-1/8}\times \Re [e^{-2\sqrt{2}(1+i)z^{1/4}+i \phi_1}]\;,
\label{firstbeha}
\ee
\be \MeijerG{3}{0}{0}{4}{-}{0,\frac{1}{2},1,-\frac{1}{2}}{z} \sim 
c_2\, z^{-1/8}\times \Re [e^{-2\sqrt{2}(1+i)z^{1/4}+i \phi_2}]
\label{secondbeha}
\ee
for large real positive $z$, $\phi_1,\phi_2$ constant real phases, and $c_1\approx c_2\approx 2.6$ (see formulas (\ref{rsexpansion}), (\ref{calcolettiparametri}) in Appendix \ref{Appendix} ). In Figs. \ref{figBLGcoefff1},\ref{figBLGcoefff2} we plot a rescaling of the components $f_1,f_2$ of the BLG Matsubara propagator, properly chosen in order to offset their asymptotic decays. Such rescalings are clear from Eqs. (\ref{firstbeha}, \ref{secondbeha}, \ref{effe1}, \ref{effe2}). In Fig. \ref{figBLGcoefffvsdistance} the same components $f_1,f_2$ are plotted without any exponential rescaling, for a fixed value of the Matsubara frequency ($\hbar \omega_0=10$ meV) and restricted to a radial range of the order of few tenths of nm.
\begin{figure}[ht]
\begin{center}
\subfigure[]{
   \includegraphics[width=0.99\columnwidth]{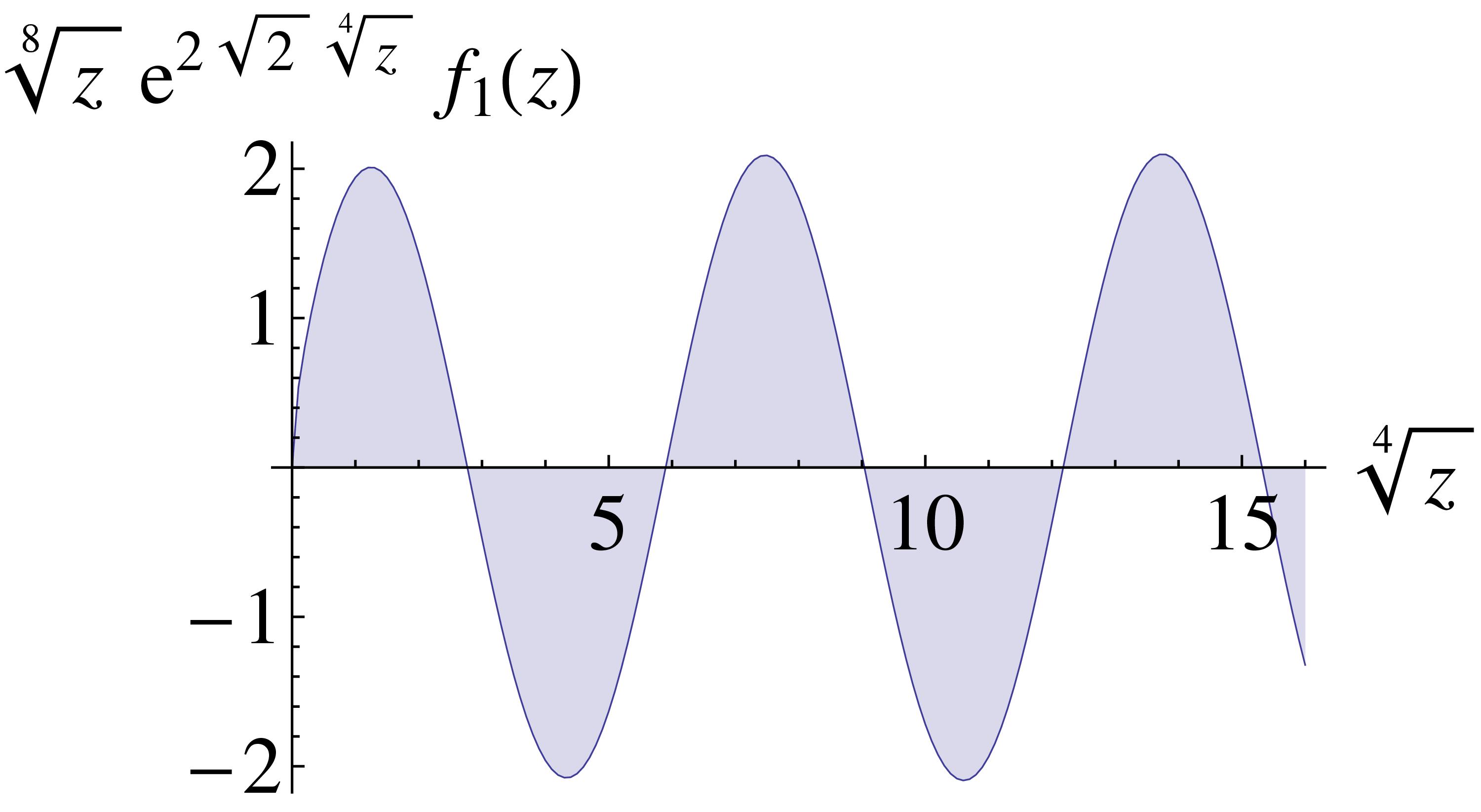}
   \label{figBLGcoefff1}
 }

 \subfigure[]{
   \includegraphics[width=0.99\columnwidth]{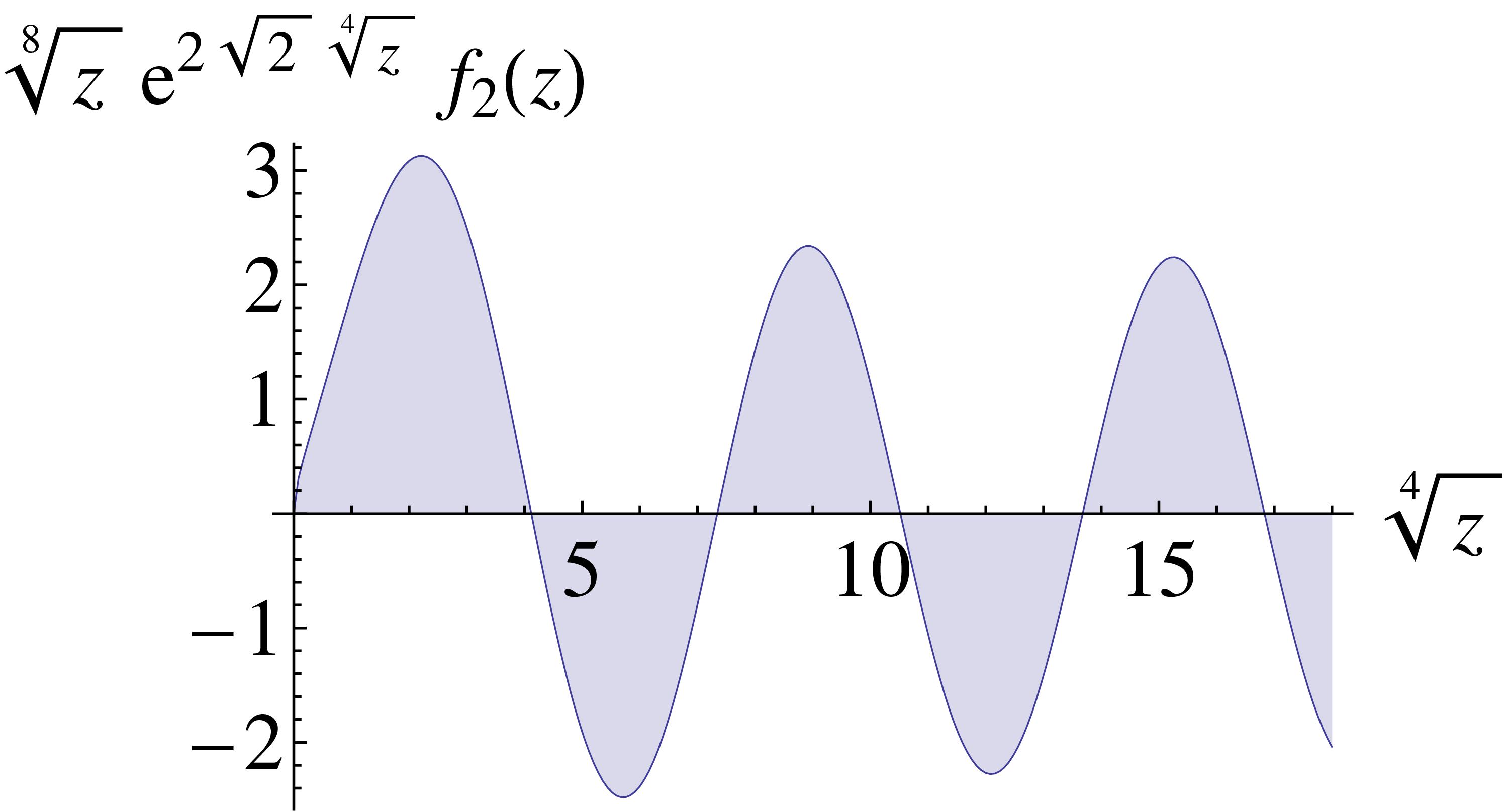}
   \label{figBLGcoefff2}
}
\subfigure[]{
  \includegraphics[width=0.49\columnwidth]{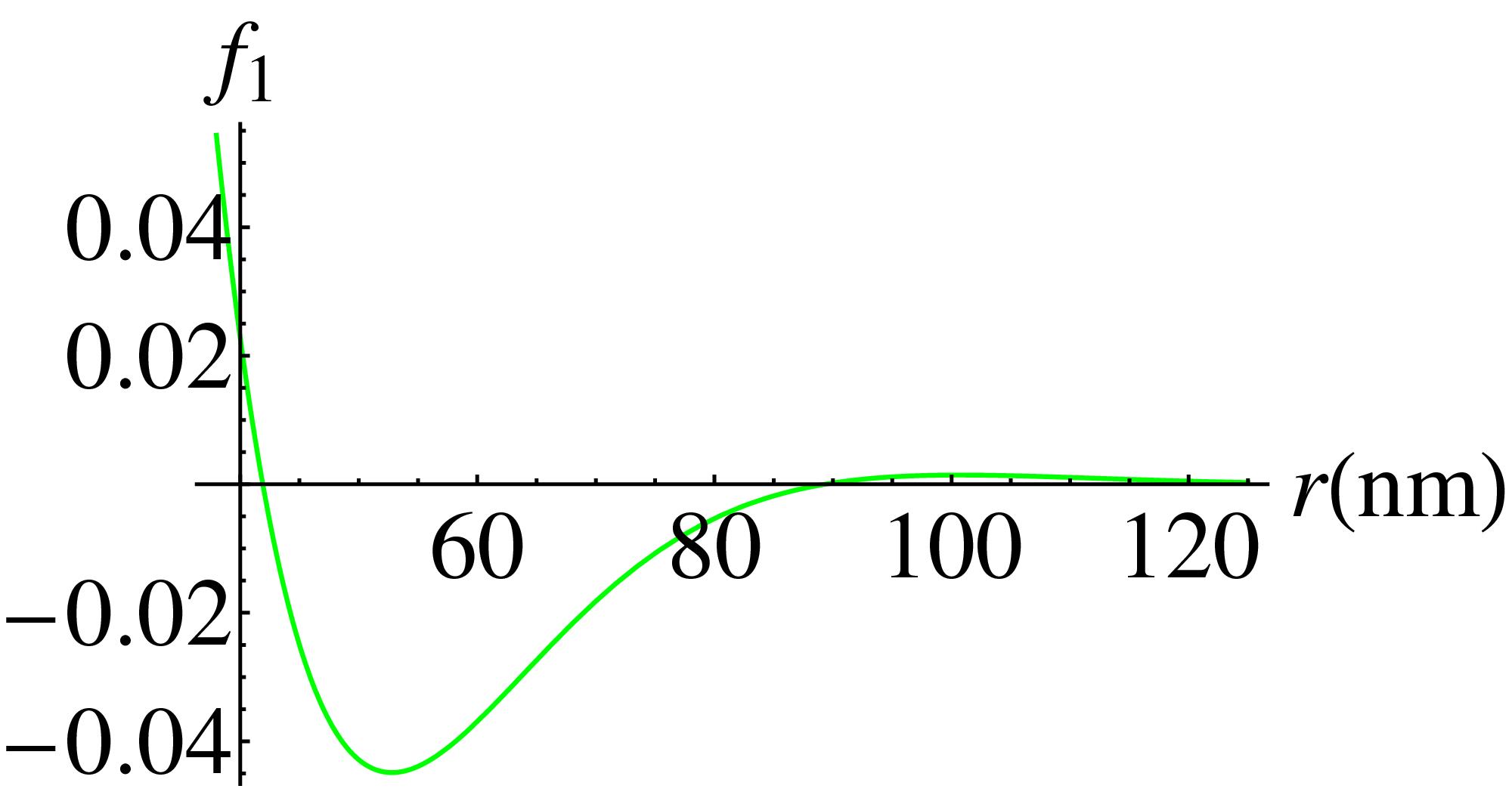}

  \includegraphics[width=0.49\columnwidth]{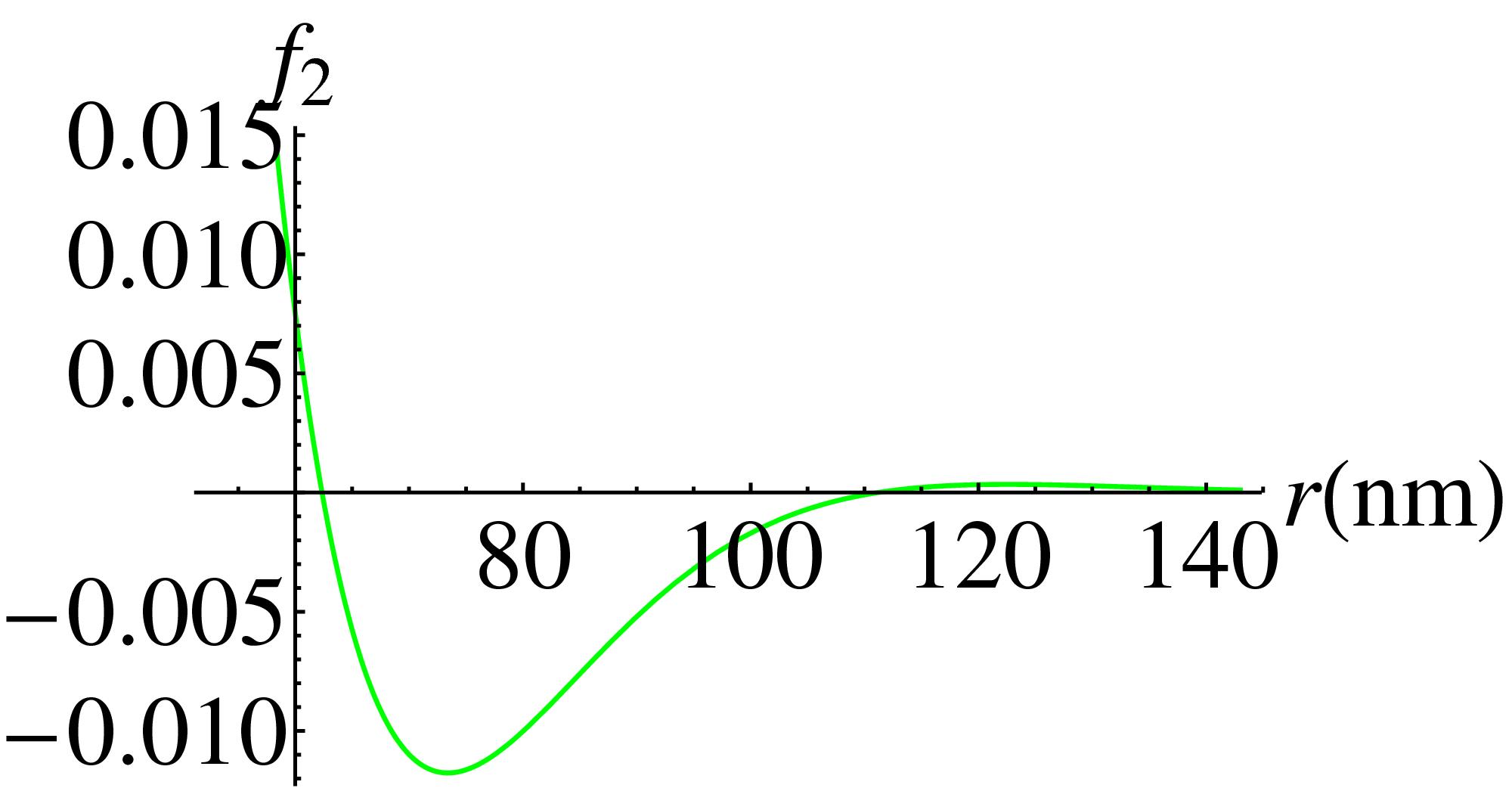}
  \label{figBLGcoefffvsdistance}
}
\end{center}
\caption{Distance-dependence of the components $f_1, f_2$ (plotted in arbitrary units) for the Matsubara (matrix) propagator of BLG for vanishing electric field ($u=0$). (a) and (b): $f_1, f_2$ are expressed in terms of the dimensionless combination $\sqrt[\leftroot{-2}\uproot{2}4]{z}\equiv r\, \sqrt[\leftroot{-2}\uproot{2}4]{m^2(u^2 + 4 \omega^2)}/4$ and rescaled by the prefactor $z^{1/8}e^{2\sqrt{2}\,z^{1/4}}$. (c) $f_1$ (left panel) and $f_2$ (right panel) are represented at Matsubara frequency $\hbar\omega_0=10$ meV as functions of the distance $r$ in units of nm.}
\label{figBLGcoeffs}
\end{figure}

At this point, for large distance $r \gg \sqrt{\hbar^2/m k_B T}$, the supercurrent looks as follows up to a constant factor, where we recall $\omega_n=(2n+1) \pi k_B T$
\begin{align}
I(\vec{r},T) \sim&
	k_B T \sum_{n=0}^{\infty}\left[ (w_n^2+\frac{u^2}{4})f_1^2(\omega_n)+f_2^2(\omega_n) \right]
\\\sim&
	k_B T \left[ (w_0^2+\frac{u^2}{4})f_1^2(\omega_0)+f_2^2(\omega_0) \right]
\\\sim&
	\frac{k_B T m^2}{\hbar^4} \left(\left[ \MeijerG{3}{0}{0}{4}{-}{0,\frac{1}{2},\frac{1}{2},0}{z}\right]^2
	+
	\left[\MeijerG{3}{0}{0}{4}{-}{0,\frac{1}{2},1,-\frac{1}{2}}{z} \right]^2\right).\label{general32}
\end{align}
Therefore, by introducing the notations
\begin{align} 
\lambda_T^{(u)}\equiv&
	\frac{h}{\sqrt{m(u^2+4 \pi^2 k_B^2 T^2)^{1/2}}}, 
\end{align}
$\lambda_T\equiv \lambda_T^{(u)}\vert_{u=0}$ being the thermal de Broglie electron wavelength, it follows that
\begin{align}
\label{superdecayzeropotasymm}
I(\vec{r},T)
	\stackrel{\left( r \gg \frac{2}{\pi}\lambda_T^{(u)}\right)}{\sim}
	-\frac{8(c_1^2+c_2^2)}{k_B T\lambda_T^4}
	\frac{\lambda_T^{(u)}}{4\pi r}
	e^{-4\pi r/\lambda_T^{(u)}}
\end{align}
and for the symmetric case $u=0$, in particular:
\begin{align}
\label{superdecayzeropotsymm}
I(\vec{r},T)\vert_{u=0}
	\stackrel{\left( r \gg \frac{2}{\pi}\lambda_T\right)}
	\sim
	-\frac{8(c_1^2+c_2^2)}{k_B T\lambda_T^4}\frac{\lambda_T}{4\pi r}
	e^{-4\pi r/\lambda_T}
\end{align}
The length $2\lambda_T/\pi$ is the reference distance scale for the validity of the approximations (\ref{superdecayzeropotasymm}) and (\ref{superdecayzeropotsymm}); this length may substantially exceed the superconducting coherence lengths of typical metals, which is the distance regime of interest in this paper, therefore either the large distance case (\ref{superdecayzeropotasymm}) and the general case (\ref{general32}) may comply with a description of the proximity effect in terms of Cooper pair propagation.

\section{Comparison with former results}\label{secV}
In Ref. \onlinecite{Gonzalez08}, the supercurrent through graphene is evaluated as the Fourier transform 
\begin{align}
I(r,T)\propto&
	\int\limits_0^\infty \frac{dk}{2\pi} \,k \, J_0(k r)\,D(\vec{k},\omega=0)\, e^{-k/k_c}\,,
\end{align}
where $k=\vert\vec{k}\vert$, $J_0$ is the Bessel function of the first kind of order zero, $k_c$ is a regularizing short distance cutoff which does not affect the behavior of the critical current in the limit of large $r$, and the Cooper pair propagator is
\begin{align}
D(\vec{k},\omega=0)=&
	i\, \text{Tr} \int\limits \frac{d \omega_q}{2 \pi} \frac{d^2q}{(2\pi)^2} G^{(a)}(\vec{q}+\vec{k},\omega_q)\,G^{(-a)}(-\vec{q},-\omega_q)\,.
\end{align}
The high-temperature (or long-distance) limit of this propagator is approximated in formula (17) of the same reference, which looks like
\begin{align}
D(\vec{k},\omega =0)\sim&
	 -\frac{\log{2}}{\pi v_F^2}k_B T-\frac{1}{16 \pi}\frac{k^2}{k_b T},\quad \;k_BT\gg v_F k\;.  
\label{formula17gonzalez} 
\end{align}
We draw attention to two points.  First, in Ref. \onlinecite{Gonzalez08} it is claimed that for very large $T$ the second term in (\ref{formula17gonzalez}) dictates the long-distance decay of the critical current, while we claim that the first addend should be evidently the dominant one, which would result therefore in a cubic (rather than quintic) long-distance decay. Second, contrary to the approach adopted in Ref. \onlinecite{Gonzalez08}, the long-distance behavior of a 2D Fourier transform (being the supercurrent, in this case) of a given function cannot be determined in general by Fourier transforming its small-momentum limit expression. A counterexample in support of this statement is given by considering the Fourier transform of the 2D Klein-Gordon propagator
\begin{align}
\int \frac{d^2k}{(\sqrt{2\pi})^2} e^{ik\cdot r} \frac{1}{k^2+m^2}=K_0(mr) \stackrel{\text{large }r}{\sim} \sqrt{\frac{\pi}{2  m r}} e^{-mr},
\label{Yukawa}
\end{align}
while instead the Fourier transform of its small-momentum approximation $m^{-2}$ is 
\begin{align}
\int \frac{d^2k}{(\sqrt{2\pi})^2} e^{ik\cdot r} \frac{1}{m^2}\propto \delta^{(2)}(\vec{r})=\frac{\delta(r)}{\pi \, r},\label{smallmomapp}
\end{align}
and e.g. even the regularizing exponential cut-off adopted in Refs. \onlinecite{Gonzalez08,Gonzalez07} would not reproduce the asymptotic exponential decay of the $\mathcal{F}$-transformed full propagator (\ref{Yukawa}):
\begin{align} 
\int \frac{d^2k}{(\sqrt{2\pi})^2} e^{ik\cdot r} e^{-k/k_c}\frac{1}{m^2}=\frac{k_c}{m^2\,\sqrt{1+ k_c^2 r^2}}\stackrel{\text{large }r}{\sim} \frac{1}{r}.\label{endcounterex}
\end{align}
The counterexample lined out in Eqs. (\ref{Yukawa}) | (\ref{endcounterex}), as well as the mismatch between the exponential (\ref{finiteTSLGdecay}) and power-law decays claimed in [Ref.~\onlinecite{Gonzalez08}, Eq. (20)], indicate that arguments are based on the specific profile of the electron dispersion relation and not on the physical conditions in the realistic situation. In fact, both examples illustrate how the long-distance decay of the supercurrent not necessarily is a univocal function of the small-momentum behavior of the electron Green function.

Physically, we expect an exponential rather than power law decay of the supercurrent as the distance significantly exceeds the thermal length, because  the latter represents the characteristic decay length of the electron propagator (compare with formula (\ref{slgpropagator})). For smaller distances ($r \ngg r_T$), the imaginary-time propagator at distance $r$ takes a significant value for several different Matsubara energies, resulting in particular in a cubic current decay when such energies are extremely close to each other w.r.t. the energy scale associated to the distance $r$ (i.e. $r \ll r_T $). Conversely, for $r\gg r_T$ the supercurrent is dominated by the lowest Matsubara energy, whose associated electron propagator features an exponential distance-decay of the propagator, due to the asymptotic decay of the Bessel functions $K_0, K_1$ w.r.t. their dimensionless arguments.  

\section{Conclusions}\label{Conclusions}
\label{secVI}
We have considered the SC proximity effect across single-layer graphene and bilayer graphene as function of the distance between the superconducting islands, for different temperatures, chemical potential (doping), and transverse electric fields. In terms of these variables, the junction behavior can be categorized into different regimes; in fact temperature, chemical potential (and, in the BLG case, the gate voltage between the two layers of BLG) set respective characteristic lengths $r_{T,\mu,u}=\{\frac{\hbar v_F}{k_BT}, \,\hbar v_F \mu^{-1},\,\hbar v_F u^{-1}$\}, for distances much shorter of each of which the Cooper pairs' tunneling remains basically unaffected by each parameter. The aforementioned various regimes for the proximity effect remain therefore defined in terms of the relation between the lengths $r,r_T,r_\mu$ (and also $r_u$ for BLG). Later on, we sum up the the most evident signatures of these regimes.

The Josephson effect was, moreover, studied in the dilute granular regime, i.e. when the distance $r$ between the SC islands is much larger than both the width $W$ and the superconducting coherence length $\xi$ of the islands. Therefore the 2D geometry of SLG/BLG is pivotal to our discussion. The supercurrent is well described in terms of Cooper pairs tunneling between the SC islands where the SLG and BLG act as a junction link. Besides to the intrinsic interest of transport properties for graphene based materials, the present study is motivated by several potential applications of proximity effect, see Sec.\,\ref{SecI}. On the other hand, this work can be seen as a theoretical starting point for the analysis of the BKT phase transition for impurities/graphene composites, where the density of SC grain impurities and other adjustable variables such as temperature, doping, transverse electric field, mechanically-induced (\textit{zig-zag} and/or \textit{armchair}) strain \cite{Alidoust11} in the system are treated as control parameters.

The vanishing density of states, characterizing graphene near its charge neutrality point, results into a particularly strong radial decay of the supercurrent. In fact, we confirm the previous result of cubic decay \cite{Gonzalez07} predicted for pristine SLG at zero temperature. This characteristics resembles the behavior of the Cooper pair propagation through a disordered normal metal. \cite{Dubos01} An analogous correspondence, between SLG junctions and disordered metal junctions, also holds for the 1D case in the opposite short junction regime.\cite{Titov06, Tworzydlo06} We point out that qualitative similarities between ballistic transport in graphene and diffusive metals also extend to the long-distance regime. In both physical systems the critical current at distances much larger than the thermal length follows an exponential decay, which supersedes the short-distance power-law decay. For undoped SLG we found an asymptotic decay $\sim r^{-2} e^{-2\pi r k_B T/ \hbar v_F}$, beyond the thermal length $r_T$ (of the
order of $\sim v_F/k_B T$).

By shifting the Fermi level away from the neutrality point, the supercurrent gets damped. This is encoded in our formulation by the introduction of a finite chemical potential, corresponding for instance to the effect of a gate-voltage or a doping procedure. For a SLG sheet, the corresponding finite density of states at the Fermi level induces a drop in the supercurrent radial decay, leading to a new regime far beyond the critical length $\hbar v_F/\mu$. The finite charge density begins to have an influence at about this length scale, with a smooth transition from the $\sim 1/r^3$ behavior, described in formula (\ref{formula11a}) and the following lines, to the faster $\sim 1/r^4$ decay at long distances, both at zero temperature.

Here, a comparison between our 2D geometry and the physics of 1D junctions (carbon nanotubes) is in order. In the case of long and narrow junctions the current can be described in terms of 1D propagation of the Cooper pairs, decaying as $\propto 1/ \mathcal{L}$ ($\mathcal{L}$ is the length of the 1D junction). \cite{Fazio96,Dolan74} Additionally, for carbon nanotubes the Coulomb interaction affects the supercurrent behavior as described in Refs. [\onlinecite{Gonzalez01,Gonzalez03}] because of a strong power-law suppression of the density of states, which turns out to be marginal for graphene.\cite{Gonzalez94,Gonzalez99}

Finally, we considered the supercurrent across bilayer graphene (firstly in the absence of inter-layer gate voltage $u$), for which we found a quadratic radial decay, $\sim 1/r^2$, at zero temperature. This characteristics reflects the quadratic low-energy dispersion of BLG band structure. For finite temperatures, we retained the asymptotic exponential distance-decay $\sim r^{-1} e^{-4\pi r/\lambda_T}$, beyond the thermal length, similarly as for SLG. The effect of a transverse electric field on the supercurrent, and its radial dependence, was considered in analytical terms. We found in (\ref{blgargument}) that $I(r,T=0)$ is proportional to $r^{-2}$ times a function of ($u r^2$) only, which generalizes the simpler decay $\sim r^{-2}$ typical of the aforementioned non-gated case $u=0$. Similarly, at finite temperature and large distance $r \gg (2/\pi) \lambda_T^{(u)}$, $\lambda_T^{(u)}\equiv h/[m^{1/2}(u^2+4\pi^2 k_B^2T^2)^{1/4}]$, the asymptotic distance-decay $\sim r^{-1} e^{-4\pi r/\lambda_T^{(u)}}$ (see (\ref{superdecayzeropotasymm})) is found to occur as a generalization of the distance-decay in absence of transverse field.

\section{Acknowledgments}\label{Acknowledgments}
We thank David Abergel for helpful discussions about numerical implementation. This work was funded by NORDITA Grant number ERC 321031-DM and the Swedish Research Council; the work of AVB was supported by US DOE E 304.

\begin{widetext}
\section{Appendix}\label{Appendix}
\subsection*{Asymptotic supercurrent through BLG}
We rewrite (\ref{effe1},\ref{effe2}) as
\be \label{compactformeffe1effe2}
f_1=\frac{m}{\sqrt{u^2 + 4\omega^2}}\,
 \MeijerG{3}{0}{0}{4}{-}{0,\frac{1}{2},\frac{1}{2},0}{z}\,,
\quad \quad
f_2=\frac{m}{2}\,
 \MeijerG{3}{0}{0}{4}{-}{0,\frac{1}{2},1,-\frac{1}{2}}{z}\,, 
\ee
where $z\equiv\frac{1}{256} m^2 r^4 (u^2 + 4 \omega^2)$. Since in both $f_1$ and $f_2$ cases there are no $\mathbf{a_p}$ parameters in the Meijer functions, the hypotheses (1.1) of Ref.  \onlinecite{Fields72} are automatically fulfilled: 
\be \label{automaticallyfulfilled} \left\{ \begin{array}{l}0\leq m \leq  q, \quad 0\leq n \leq  p,  \\
a_j-b_k \neq \text{a positive integer, } j=1,\cdots p;\,k=1,\cdots , q,\\
 a_j-a_k \neq \text{an integer, } j,k=1,\cdots p;\,j \neq k\,.\end{array} \right.
 \ee
We are in the $q>p$ case, and define $\nu\equiv q-p=4$,\, $\mu\equiv q-m-n=1$. Theorem 3 of Ref. \onlinecite{Fields72} guarantees that, provided $\nu\geq 1$, and called $(r,s)$ each pair of integer numbers fulfilling
\be  \begin{array}{l} \vert \arg z+\pi(\mu+1-2 r)\vert < \pi(\nu/2+1), \\
\vert \arg z+\pi(\mu+2-2s-2h)\vert < \pi(\nu+\min(1,\nu/2)), \quad h=1,\cdots,\nu \,,\end{array}
\ee
if the sector 
\be \label{ipotesidelteorema3} S_{r,s}: \begin{array}{l} \pi(\nu-\mu-2+\max[2r-\frac{3 \nu}{2}, 2s-\min(1,\frac{\nu}{2})])< \arg z \\ \arg z < \pi(\frac{\nu}{2}-\mu+\min[2r,\frac{\nu}{2}+2s+\min(1,\frac{\nu}{2})])  \end{array}
\ee
is not empty, then suitable constants $C_i(r,s), D_i(r,s)$ exist, such that:
\be \label{rsexpansion} G^{m,n}_{p,q}=\sum_{j=1}^{p} C_j(r,s)\,L_j(z\,e^{i\pi(\mu+1-2 r)})+\sum_{h=1}^{\nu} D_h(r,s)\,G(z\,e^{i\pi(\mu+2-2 s-2 h)})\,;
\ee
this expansion is referred to as the (r,s) expansion for $G^{m,n}_{p,q}(z)$, and $L_j$ and $G$ stands for the functions defined below:
\be \nonumber
L_j(w)\equiv \MeijerG{q}{1}{p}{q}{a_j,a_1\cdots ,a_{j-1},a_{j+1},\cdots,a_p}{b_1,\cdots, b_q}{w}\;
\sim \frac{ \Gamma(1+b_Q-a_j)}{ \Gamma(1+a_P-a_j)} \,w^{-1+a_j}\;\textstyle\pFq{q+1}{p}{1,,1+b_Q-a_j}{1+a_P-a_j}{-\frac{1}{w}}
\ee
\be 
j=1,\cdots , p, \quad w \rightarrow \infty, \quad \vert \arg w \vert <\pi(\nu /2 +1)\,,
\ee
where the following notations are intended $$\Gamma_n(c_P-t)\equiv \prod\limits_{k=n+1}^{p}\Gamma(c_k-t), \quad \Gamma(c_M-t)\equiv\Gamma_0(c_M-t), \quad \textstyle\pFq{p}{q}{a_P}{b_Q}{z}\equiv \sum\limits_{k=0}^{\infty}\frac{\Gamma(a_P+k)\Gamma(b_Q)}{\Gamma{b_Q+k}\Gamma{a_P}}\;; $$
\be \nonumber
G(w)\equiv \MeijerG{q}{0}{p}{q}{a_1,\cdots ,a_p}{b_1,\cdots, b_q}{w} \sim\left(\frac{(2\pi)^{\nu -1}}{\nu}\right)^{1/2}\,e^{-\nu\,w^{1/\nu}}\;\sum_{j=0}^{\infty}K_j\,
w^{\gamma - \frac{j}{\nu}}\,, 
\ee
\be 
 w \rightarrow \infty, \quad \vert \arg w \vert <\pi(\nu +\min(1,\nu/2))\;,
\ee
where
\be \nonumber
\nu\gamma=\frac{1-\nu}{2}+B_1-A_1,\quad\quad  K_0=1 \;,
\ee
\be \label{calcolettiparametri}
K_1=A_2-B_2+\frac{B_1-A_1}{2\nu}[\nu(A_1+B_1)+A_1-B_1]+\frac{1-\nu^2}{24\nu}\;,
\ee
\be \nonumber
\prod_{j=1}^{p}(x+a_j)=\sum_{j=0}^{p}A_j\,x^{p-j},\quad\quad  
\prod_{j=1}^{q}(x+b_j)=\sum_{j=0}^{q}B_j\,x^{q-j}\,,
\ee
and the remaining $K_j$ are polynomials in $A_j,B_j$ independent of $w$. In both cases (\ref{compactformeffe1effe2}), one can easily check the occurrence of the hypotheses (\ref{automaticallyfulfilled},\ref{ipotesidelteorema3}) required for the application of Theorem 3 of Ref. \onlinecite{Fields72}, therefore the expansion (\ref{rsexpansion}) holds.
\end{widetext}

\end{document}